\documentclass[10pt]{article} 
\usepackage[preprint]{rlc}

\usepackage{amssymb}            
\usepackage{mathtools}          
\usepackage{mathrsfs}           
\mathtoolsset{showonlyrefs}     
\usepackage{graphicx}           
\usepackage{subcaption}         
\usepackage[space]{grffile}     
\usepackage{url}             
\usepackage{xurl}
\usepackage{rotating}

\newtoggle{final}
\togglefalse{final} 

\def\email{\small\ttfamily}
\title{\centering\raisebox{-0.2em}{\includegraphics[height=1em]{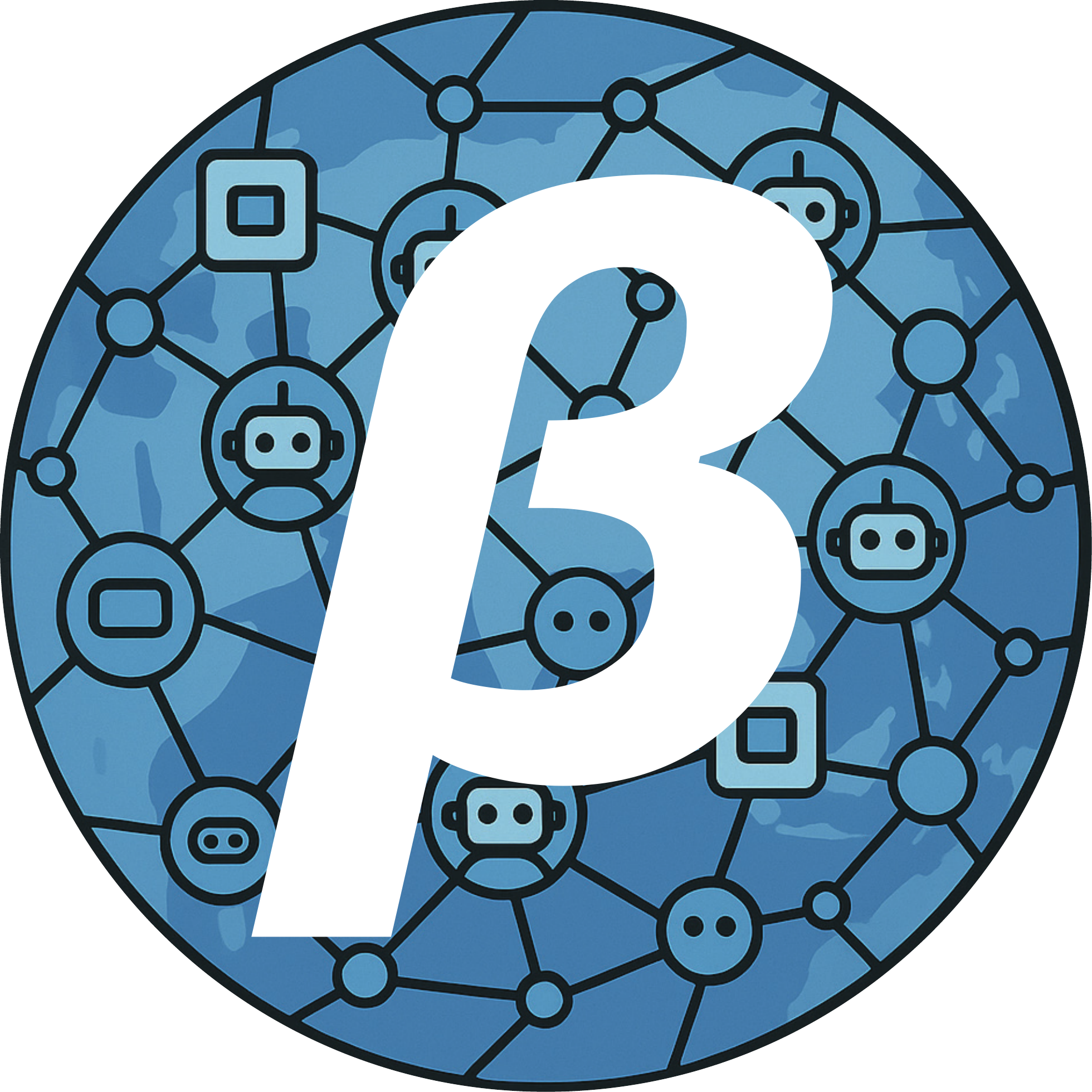}} BetaWeb: Towards a Blockchain-enabled\\Trustworthy Agentic Web}

\author{
\parbox{\textwidth}{
\centering
Zihan Guo$^{1,2}$, Yuanjian Zhou$^{1}$, Chenyi Wang$^{1,3}$, Linlin You$^{2}$,
\\
Minjie Bian$^{4}$, Weinan Zhang$^{1,5*}$
}
\vspace{0.5em} 
\\
$^{1}$ Shanghai Innovation Institute\quad$^{2}$ Sun Yat-sen University
\\
$^{3}$ Zhejiang University\quad$^{4}$ Shanghai Data Group Co., Ltd
\\
$^{5}$ Shanghai Jiao Tong University
\\
\email{guozh29@mail2.sysu.edu.cn, wnzhang@sjtu.edu.cn}
}

\begin{document}

\maketitle

\begin{abstract}

The rapid development of large language models (LLMs) has significantly propelled the development of artificial intelligence (AI) agents, which are increasingly evolving into diverse autonomous entities, advancing the LLM-based multi-agent systems (LaMAS). However, current agentic ecosystems remain fragmented and closed. Establishing an interconnected and scalable paradigm for Agentic AI has become a critical prerequisite.
Although Agentic Web proposes an open architecture to break the ecosystem barriers, its implementation still faces core challenges such as privacy protection, data management, and value measurement. Existing centralized or semi-centralized paradigms suffer from inherent limitations, making them inadequate for supporting large-scale, heterogeneous, and cross-domain autonomous interactions.
To address these challenges, this paper introduces the blockchain-enabled trustworthy Agentic Web (BetaWeb). By leveraging the inherent strengths of blockchain, BetaWeb not only offers a trustworthy and scalable infrastructure for LaMAS but also has the potential to advance the Web paradigm from Web3 (centered on data ownership) towards Web3.5, which emphasizes ownership of agent capabilities and the monetization of intelligence.
Beyond a systematic examination of the BetaWeb framework, this paper presents a five-stage evolutionary roadmap, outlining the path of LaMAS from passive execution to advanced collaboration and autonomous governance. We also conduct a comparative analysis of existing products and discuss key challenges of BetaWeb from multiple perspectives. Ultimately, we argue that deep integration between blockchain and LaMAS can lay the foundation for a resilient, trustworthy, and sustainably incentivized digital ecosystem. A summary of the enabling technologies for each stage is available at \url{https://github.com/MatZaharia/BetaWeb}.

\vspace{10pt}
\end{abstract}

\section{Introduction}

With the rapid and transformative advances in large language models (LLM), artificial intelligence (AI) agents are evolving from early rule-based and passively executed units into autonomous entities endowed with end-to-end capabilities encompassing perception, learning, decision-making, and execution \citep{wang2024survey, guo2024large}. The system architectures for agents are also shifting from single-node deployments to distributed and multi-agent collaborative frameworks \citep{gao2025single}. Through shared perception, task planning, and interactive cooperation among multiple agents, these systems can accomplish increasingly complex and dynamic objectives. It is an evolution that profoundly reshapes the design principles and application paradigms, driving the advancement towards LLM-based multi-agent systems (LaMAS) \citep{yang2024llm}.

\begin{figure}[htb]
    \centering
    \includegraphics[width=\textwidth]{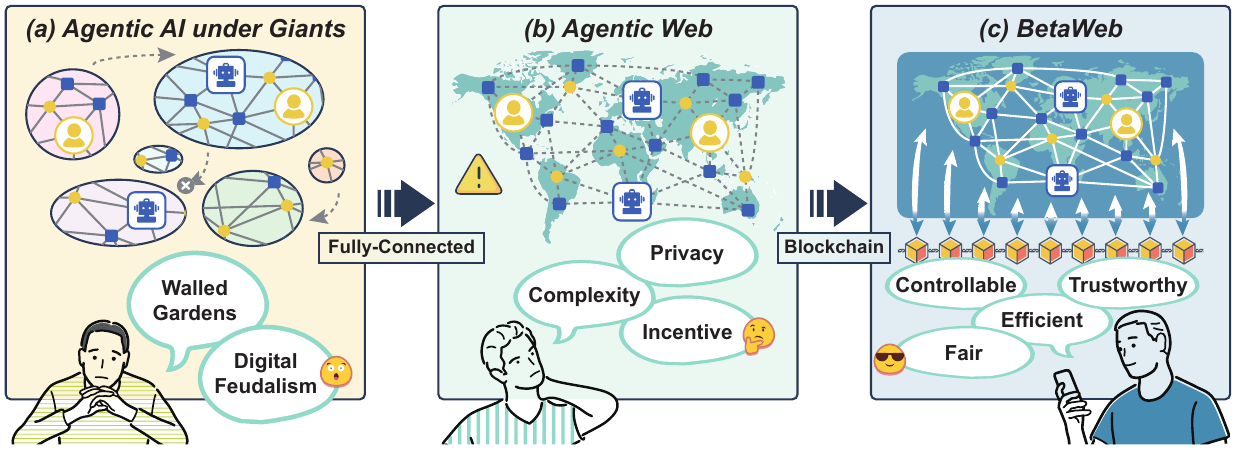}
    \caption{Schematic illustration of the evolution from siloed and platform-controlled agentic AI ecosystems to BetaWeb. (a) Current stage dominated by a few major giants, where users and agents are tightly connected within the ``walled garden'' of each platform, with only weak inter-platform links, reflecting digital feudalism concerns. (b) Conceptual open Agentic Web, where users and agents are globally connected, but face challenges such as privacy protection, coordination complexity, and incentive alignment. (c) Vision of BetaWeb, where decentralized infrastructure supports trustworthy, controllable, efficient, and fair interactions among globally connected users and agents.}
    \label{fig:web-evolution}
\end{figure}

Although current LaMAS demonstrate remarkable performance in controlled environments, their potential remains constrained by a siloed reality \citep{petrova2025semantic}. As shown in Figure \ref{fig:web-evolution}a, the existing ecosystems of Agentic AI \citep{sapkota2025ai} are still predominantly platform-centric under various giants, such as Google Agentspace \citep{googleagentspace} and AWS Marketplace \citep{awsmarketplace}, where users and the growing agents operate largely within the walled gardens controlled by centralized entities \citep{scheuerman2024walled}. Such a form of digital feudalism concentrates data ownership and interaction protocols in the hands of platform gatekeepers, thereby limiting system transparency and user rights, including practices like big data price discrimination \citep{pandey2021disparate}, algorithmic content manipulation \citep{ienca2023artificial}, and opaque user data sharing policies \citep{bietti2025data}. Meanwhile, individual capabilities in the traditional Internet have been constrained by limited operational tools and fragmented interaction interfaces, which are now being lifted by unleashing unprecedented potential for large-scale autonomy and collaboration with agents, fueling an urgent demand for open and interoperable digital environments \citep{wang2025internet}.

Especially as autonomous agents increasingly collaborate across organizational boundaries, heterogeneous platforms, and diverse application scenarios, the construction of an open, interoperable, and scalable coordination layer has become a critical task. This vision strongly aligns with the emerging concept of Agentic Web\footnote{Agentic AI, Agentic Web, and LaMAS are often easily confused. Conceptually, \textbf{Agentic AI} represents a foundational architectural innovation and a leap forward in intelligent organizational forms, reflecting a new understanding of agent behavior and their interactions, with an emphasis on autonomy and collaboration. \textbf{Agentic Web}, on the other hand, is a network architecture designed for agents, aimed at supporting them in accomplishing tasks and goals through communication, collaboration, and resource sharing. In contrast, \textbf{LaMAS} are systematic tools that, through specific system-level designs, support the interaction and autonomous behavior of multiple agents in practical applications. \textbf{In short, Agentic AI focuses on a paradigm shift at the ecological level, LaMAS provide the specific systems to achieve this shift, and the Agentic Web offers an Internet infrastructure for agent collaboration within or across systems.}} \citep{yang2025agentic}, which is a fully-connected architecture driven by agent participation as illustrated in Figure \ref{fig:web-evolution}b. Its goal is to overcome the collaboration barriers of closed ecosystems, enabling a new paradigm of the machine-to-machine economy. Agentic Web also support large-scale knowledge integration and adaptive cross-domain problem-solving, and facilitate social behaviors as well as co-creation of value across organizations and platforms.

The transition to the Agentic Web not only represents a technological evolution but also signifies an inevitable transformation of the digital ecosystem. Despite the strong appeal of this vision, its realization faces a series of urgent challenges that current digital infrastructures struggle to address. As agents increasingly transcend single-platform boundaries to operate across domains, critical issues such as trust establishment, verifiable identities, transparent interaction records, and secure coordination become ever more prominent \citep{raza2025trism,sharma2025collaborative,yin2024research}. Without a reliable and trustworthy foundational architecture, large-scale agent collaboration risks being hindered by fragmented standards, opaque governance, and susceptibility to manipulation, potentially resulting in uncontrollable outcomes \citep{deng2025ai,puthal2025shadow,sharma2025collaborative}. Specifically, the key challenges can be summarized into the following three areas:

\begin{enumerate}
    \item Privacy Protection and Risk Control: Openness brings collaboration opportunities but also amplifies challenges related to privacy breaches, unclear legal responsibilities, and risk management. Given that agents operate at higher speeds and scales compared to human users, the impact of errors or malicious actions may be exponentially magnified. Therefore, it is essential to ensure agent controllability and achieve traceability and immutability of their behaviors \citep{narajala2025securing,krishnan2025ai}.
    \item Complexity of Coordination and Data Management: Under the prospect of agents infinitely scaling and working continuously, the volume and frequency of interactions will far exceed the current level of information flow on the Internet, substantially increasing the complexity of system coordination, data consistency, and resource scheduling. This necessitates the construction of trust and coordination mechanisms to ensure synchronization and verifiability in cross-domain agent collaboration \citep{zhang2024scalable,qian2024scaling,pan2024very}. 
    \item Value Measurement and Individual Incentives: With the shift in the digital paradigm, agents will serve not only as data consumers but also as providers of cognitive services and computing resources. There is an urgent need to establish verifiable and auditable contribution assessment mechanisms, upon which fair and transparent incentive and reward distribution can be based, thereby fostering a healthy and sustainable agent ecosystem economy \citep{chaffer2025can,ricci2024cognitive}.
\end{enumerate}

These challenges collectively underscore the necessity of building a trustworthy Agentic Web. While several common paradigms can partially address the needs for trust, coordination, and governance in ecosystem construction, their inherent limitations make them ill-suited for future open, heterogeneous, and cross-domain scenarios. For example, 1) traditional public key infrastructure \citep{dumitrescu2024failures} offers high robustness in identity authentication and encrypted communication, yet it was never designed to support universal consensus in heterogeneous networks, nor can it natively enable tamper-proof and multi-party state sharing. 2) Consortium-based \citep{bai2021public} or federated modes \citep{ksystra2022towards} distribute decision-making power among multiple stakeholders, which can mitigate single-point control risks within a defined scope. However, they still rely on pre-established trust relationships and remain susceptible to collusion and conspiracy, rendering them unsuitable for open, large-scale, cross-domain agent collaboration. 3) Trusted execution environments \citep{geppert2022trusted} and other hardware-based trust anchors \citep{sardar2023confidential} provide tamper-resistant computation and isolated execution environments, but their reliance on closed, proprietary ecosystems limits scalability and deprives them of the openness and interoperability required for an Internet-scale, agent-driven economy.

The limitations indicate that existing centralized or semi-centralized paradigms cannot fully meet the demands of a truly open, interoperable, and high-throughput system, particularly in scenarios where agents autonomously transact, require verifiable transaction records, and must support auditable contribution-based incentive mechanisms \citep{wang2022blockchain,karim2025ai}. Intuitively, blockchain technology offers a compelling solution for this purpose. Its native integration of decentralized consensus, immutable record-keeping, and programmable trust logic not only enables verifiable agent identities and traceable interaction histories but also ensures the distributed nature of trust and the inherent resilience of the system, making blockchain an ideal foundation for trustworthy Agentic Web. Specifically, the advantages of blockchain can be directly mapped to the core challenges:

\begin{enumerate}
    \item The combination of immutable on-chain records with privacy-preserving cryptographic techniques, such as zero-knowledge proofs, enables interaction traceability while safeguarding participant privacy, significantly reducing legal and operational risks in open environments \citep{zhou2024leveraging}.
    \item Consensus protocols and distributed ledger mechanisms in blockchain can establish a globally consistent and verifiable system state within a heterogeneous and cross-domain Agentic Web, thereby effectively addressing the high complexity of cross-domain collaboration and data synchronization \citep{bellaj2024drawing}.
    \item Programmable smart contracts provide a native pathway for verifiable value attribution and for automated and transparent incentive distribution. This ensures that the contributions of agents (whether in computation, knowledge provision, or decision support) can be rewarded fairly in an auditable manner \citep{xu2024model}.
\end{enumerate}

In this paper, we propose the blockchain-enabled trustworthy Agentic Web (BetaWeb, $\beta$-Web). In BetaWeb, blockchain not only resolves critical technical challenges but also, when integrated with LaMAS system, lays the groundwork for a resilient, self-sustaining, and incentive-aligned machine-to-machine economy. This integration provides a solid basis for future autonomous and scalable digital ecosystems. Beyond traditional Web3, which primarily focuses on data ownership, BetaWeb extends to the realms of intelligent ownership and a capability economy, upgrading the paradigm from ``owning and monetizing data'' to ``owning and monetizing intelligence'', marking the transition towards Web3.5.

Therefore, it is essential to re-examine LaMAS from the emerging perspective of BetaWeb, especially analyzing how foundational technologies and modular capabilities of blockchain can serve as an enabling underlying infrastructure to support open collaboration and trustworthy operation. Against this backdrop, core components and involved parties should be redefined and redesigned from an agent-centric manner, particularly in trust negotiation, autonomous interaction, and incentive alignment. Although both Agentic AI and blockchain are advancing rapidly and demonstrate clear synergies, there is still a lack of systematic research that maps and critically analyzes their deep integration. Bridging this research gap is essential for understanding and shaping BetaWeb with the evolutionary trajectory of future ecosystems, which is an endeavor that constitutes the central objective of this paper.

In summary, the main contributions and structure of this paper are as follows. In Section 2, we examine the transformative impacts that BetaWeb may bring when deployed at scale, with a particular focus on their potential to reshape value-exchange mechanisms and relations of production. In Section 3, we present a five-stage evolutionary roadmap towards a fully autonomous and decentralized ecosystem for Agentic AI, outlining its characteristics, key enabling technologies, and major milestones along the way. In Section 4, we analyze existing products that integrate blockchain with LaMAS, situating them within the proposed evolutionary roadmap to identify current gaps and opportunities for advancement. In Section 5, we describe the key challenges and open problems from perspectives of techniques, ecosystems, and society. In Section 6, we conclude the paper by summarizing the main arguments.

\vspace{15pt}


\section{Blockchain-enabled Trustworthy Agentic Web}

\subsection{General Framework}

Although we have analyzed the paradigm integration and technical feasibility of BetaWeb at a conceptual level, their practical implementation still demands a well-defined and systematic architecture. Specifically, it is necessary to clearly articulate how human users, agents, and the underlying agentic workflows within a decentralized, trustless, and auditable environment. Without a unified framework, the integration process risks becoming fragmented, serving only narrow application scenarios and failing to support a sustainable and scalable ecosystem \citep{zou2025blocka2a}.

\begin{figure}[htb]
    \centering
    \includegraphics[width=\textwidth]{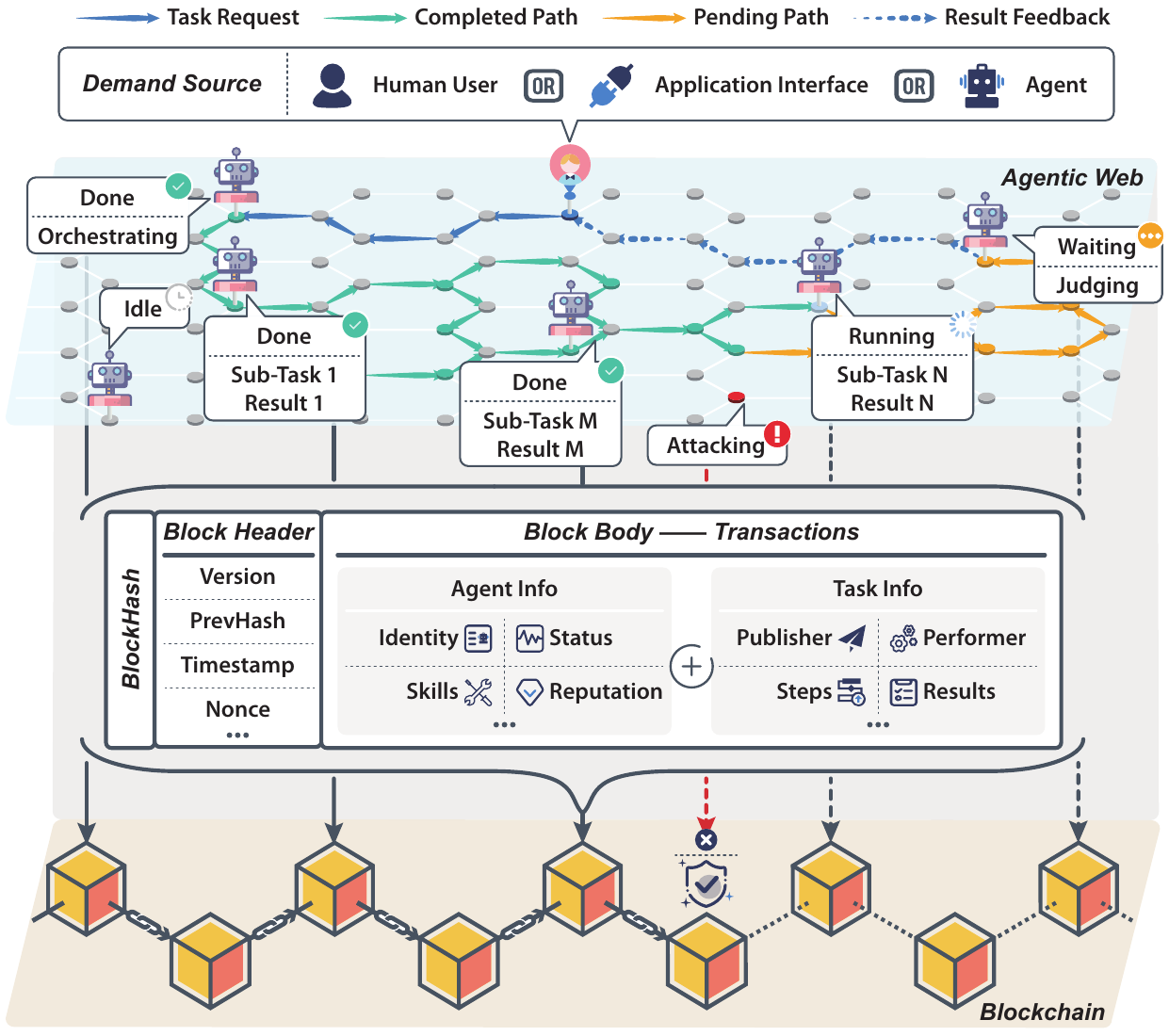}
    \caption{General framework of BetaWeb. All interactions are abstracted as task procedures from request to feedback, where the demand source (either a human user, an application interface, or another agent) invokes tasks that may be decomposed into multiple sub-tasks, executed autonomously by agents. The blockchain serves as the trusted substrate for the full lifecycle management of both agents and tasks, ensuring verifiable identity, immutable records, and transparent governance.}
    \label{fig:general-framework}
\end{figure}

To address this issue, we outline a general framework, illustrated in Figure \ref{fig:general-framework}, which abstracts all interactions into standardized task procedures, including the request, execution, and feedback. Therefore, the operational logic of LaMAS within BetaWeb is redefined. Regardless of whether an request originates from a human user, an API interface, or another agent, the associated actions will be formally recorded as transactions related to the task in the blockchain. This task can be decomposed into multiple steps, namely sub-tasks, which are autonomously orchestrated and executed by agents \citep{sun2025data}. Notably, the entire lifecycle of all agents and tasks is anchored into the blockchain for recording and governing to enable verifiable and tamper-resistant process control.

Evidently, by embedding a decentralized consensus infrastructure in BetaWeb, the framework ensures system-level verifiability of agents and tasks, with both identities and interaction processes cryptographically verifiable \citep{islam2024mrl}. At the same time, all related records are immutable, effectively preventing data tampering, deletion, and other erroneous or malicious operations \citep{karim2025ai}. Furthermore, agent behaviors are fully auditable, enabling precise attribution of contributions and responsibilities, thereby facilitating transparent accountability mechanisms and equitable incentive distribution \citep{sami2024learnchain}. 

Overall, this design not only provides a sustainable, scalable, and governable foundational support for building large-scale, high-concurrency, and loosely coupled BetaWeb, but also fosters the emergence of an adaptive and iteratively upgradable ecosystem, laying a solid foundation for a viable pathway towards realizing trustworthy, controllable, efficient, and fair Agentic AI \citep{sapkota2025ai}.

\subsection{Key Involved Parties}

The key involved parties in BetaWeb can be categorized into three major dimensions, corresponding to profound transformations at the levels of users, agents, and agentic workflows \citep{caetano2025agentic}.

At the user level, there is a fundamental shift in user role. Under traditional paradigms, users are required to directly engage in specific operations and decision-making processes \citep{xu2023transitioning}. In BetaWeb, however, users primarily focus on defining goals and requirements while delegating execution to autonomous agents. This transition liberates users from tedious task execution, enabling an identity shift from operator to the consumer, manager and owner of the agents \citep{leonardi2025homo}. Blockchain ensures that user identities, permissions, and actions are verifiable. This not only significantly enhances user experience and reduces human effort and operational complexity but also facilitates safer and more efficient cognitive interactions and value exchanges.

At the agent level, agents evolve from simple tools into autonomous entities endowed with perception, decision-making, learning, and execution capabilities \citep{wang2024survey}. Leveraging BetaWeb, agents are able to collaborate across domains and interact dynamically, completing complex tasks involving multiple participants \citep{han2024llm} while continuously acquiring knowledge \citep{wang2024benchmark}. Under trustworthy and controllable conditions, their autonomy and self-governance are substantially enhanced, enabling proactive responses to environmental and task changes, thereby advancing the overall intelligence of the system and driving it towards a truly autonomous ecosystem for Agentic AI \citep{hu2025trustless}.

At the agentic workflows level, it achieves high degrees of integration, automation, and systematization. Traditional agentic workflows rely heavily on manual intervention, which are characterized by static, linear, and isolated processes. They are replaced by a dynamic, closed-loop, and adaptive intelligent ecosystem \citep{yu2025survey}. Within BetaWeb, agentic workflows can be autonomously completed with self-optimization, enabling effective handling of complex tasks and self-recovery from anomalies \citep{10849561}. The entire workflow remains highly transparent and traceable, supporting automated execution driven by smart contracts as well as dynamic adjustments through autonomous governance, thereby constructing a highly autonomous and elastically scalable environment.

\subsection{Core System Modules}

Beyond general functionalities, the core system modules in BetaWeb can be summarized into two layers. The upper layer consists of business modules responsible for specific system interactions, focusing on the full lifecycle management and execution of tasks to ensure transparency and controllability from initiation to completion. The lower support layer ensures overall system management and governance, concentrating on agent management and rule enforcement, thereby guaranteeing the rational economic incentives, effective operation of autonomous mechanisms, and controllability of agents throughout their lifecycle. These two layers complement each other to jointly build a sustainable, secure, and efficient ecosystem for future BetaWeb.

Within the upper business modules, the first core component is task management. This encompasses the entire lifecycle of tasks, including request parsing, process orchestration, result verification, and incentive settlement \citep{lin2024mao}. The immutable ledger and smart contract in blockchain render the task management process highly transparent and trustworthy, ensuring every step is traceable and thereby enhancing the security and effectiveness of task execution.

The second business module is for task execution, responsible for the actual execution of all tasks or sub-tasks. Beyond the fundamental operation of agents, this module involves container scheduling \citep{xiao2024deepcts}, hybrid on-chain and off-chain data storage \citep{tsang2024chain}, invocation of various tools \citep{masterman2024landscape}, and generation of proofs \citep{vcapko2022state}. Task execution relies not only on the coordinated use of distributed computing resources but also leverages blockchain to ensure verifiability and auditability of the execution process, guaranteeing the authenticity and integrity of all outcomes and promoting efficient completion of complex tasks.

In the lower support modules, the agent management module is responsible for the entire  lifecycle of agents, which includes supervision and control of their identity, capability, reputation, and account. By recording agent behaviors immutably, combined with decentralized autonomous organizations (DAO) \citep{bonnet2024decentralized} and dynamic incentive mechanisms \citep{han2022can}, this module significantly enhances agent autonomy and reliability, driving the system towards higher trustworthiness and more efficient collaboration.

Finally, the rule management module focuses on the design, administration, and on-chain governance of smart contracts, encompassing economic models, reward and penalty mechanisms, and dynamic parameter adjustments \citep{zutshi2021value,10634828}. This module ensures the transparency, fairness, and adaptability of system rules, supporting multi-stakeholder co-governance and enabling continuous optimization based on real-world operational feedback, thereby maintaining ecosystem stability and sustainable development.

\section{Five-stage Evolutionary Roadmap}

\begin{table}[htb]
\centering
\renewcommand{\arraystretch}{2}
\caption{Overview of the five-stage evolutionary roadmap.}
\resizebox{\textwidth}{!}{
\begin{tabular}{lll}
\hline
\textbf{Stage} & \textbf{Name} & \textbf{Explanation} \\
\hline
Stage 1 (S1) & Isolated Silos & Designed LaMAS under human control with siloed governance \\
Stage 2 (S2) & Pilot Decentralization & Decentralized LaMAS still human-led with limited agentic workflows\\
Stage 3 (S3) & Assisted Execution & Agent-assisted LaMAS freeing human labor \\
Stage 4 (S4) & Hybrid Governance & Co-governed LaMAS relieving human cognitive burdens \\
Stage 5 (S5) & Full Autonomy & Autonomous LaMAS with humans setting only the overarching direction \\
\hline
\end{tabular}
}
\label{tab:stage-explanation}
\end{table}

\begin{figure}[htb]
    \centering
    \includegraphics[width=\textwidth]{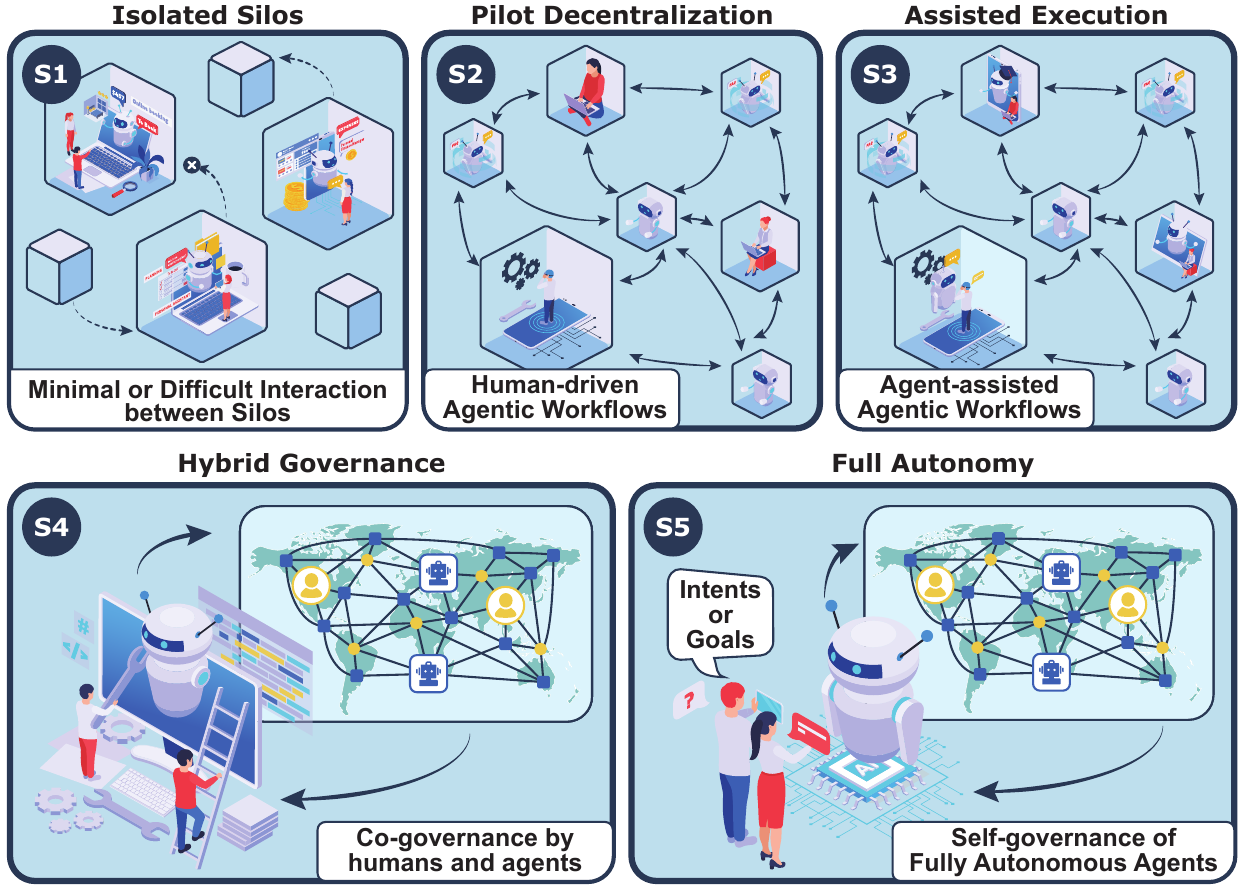}
    \caption{Five-stage evolution diagram of BetaWeb. Stage 1 (Isolated Silos) shows independent systems where humans drive all tasks, and agents are confined within their platforms with minimal outward interaction. Stage 2 (Pilot Decentralization) introduces cross-platform collaboration, but agentic workflows are completed under human supervision. Stage 3 (Assisted Execution) involves agents to undertake specialized duties, reducing the workload on humans. Stage 4 (Hybrid Governance) depicts large-scale distributed collaboration, where agents participate in governance while humans focus on high-value decisions. Stage 5 (Full Autonomy) represents a fully autonomous system where agents operate globally with end-to-end self-management without human intervention, requiring only the presentation of intents or goals.}
    \label{fig:five-stage}
\end{figure}

\begin{table}[htb]
\centering
\renewcommand{\arraystretch}{2}
\caption{Comparison of key involved parties across the five-stage evolutionary roadmap.}
\resizebox{\textwidth}{!}{
\begin{tabular}{p{3cm}p{5cm}p{5cm}p{5cm}}
\hline
\textbf{Stage} & \textbf{Human Users} & \textbf{Agent Capabilities} & \textbf{Agentic Workflows} \\
\hline
Isolated Silos &
Deeply involved in every step &
Restricted to specific environments &
Closed processes within individual systems \\
Pilot \linebreak Decentralization &
Lead the entire task process according to human cognition and judgment &
General-purpose capabilities with single-step command execution &
Decentralization assisted by hard-coded static contracts \\
Assisted Execution &
Reduced day-to-day execution workload &
Internalized domain-specific capabilities &
Rules upgradable by human intervention \\
Hybrid Governance &
Focus on essential decision-making and governance strategy &
Possess self-evolution abilities, capable of continuous improvement in specific domains &
Agent-assisted governance with human-machine collaborative decision-making \\
Full Autonomy &
Provide only high-level requirements, fully freed from operations &
Fully autonomous, capable of performing social behaviors &
Fully autonomous agentic governance with no human intervention \\
\hline
\end{tabular}
}
\label{tab:stage-parties}
\end{table}

\begin{table}[htb]
\centering
\renewcommand{\arraystretch}{2}
\caption{Comparison of core system modules across the five-stage evolutionary roadmap (cont.).}
\resizebox{\textwidth}{!}{
\begin{tabular}{p{3cm}p{3.8cm}p{3.8cm}p{3.8cm}p{3.8cm}}
\hline
\textbf{Stage} & \textbf{Task Management} & \textbf{Task Execution} & \textbf{Agent Management} & \textbf{Rule Management} \\
\hline
Isolated Silos &
Human-centric process without cross-entity collaboration &
Completed within isolated systems &
Generally non-existent or extremely basic &
Static rules maintained manually \\
Pilot \linebreak Decentralization &
Fixed workflows analyzed, defined, and managed by humans &
Simply invoke agents, primarily relying on fixed functions or APIs &
Simple identity with basic information &
Immutable smart contracts manually written and deployed \\
Assisted Execution &
Agents assist in request parsing, workflow decomposition, and other upstream tasks &
Agents lead subtask execution, with capabilities modularized into pluggable components &
Capability evaluation, incentive mechanisms, and other governance tools are introduced &
Agents generate smart contracts based on natural language requests from humans \\
Hybrid Governance &
Agents assist with downstream tasks such as result verification, but humans intervene when necessary &
Proprietary-capability agents undergo continuous iterative upgrades to enhance adaptability &
Controllable agents with mechanisms for behavior monitoring, governance penalties, and forced shutdown &
Agents autonomously generate upgrade proposals based on system monitoring, subject to human approval \\
Full Autonomy &
Agents manage the entire task lifecycle autonomously, forming an end-to-end closed loop without human involvement &
Agents possess autonomous upgrading and task-space exploration capabilities, enabling self-innovation and adaptive expansion &
Agents autonomously discover, recruit, or remove members, forming a DAO composed entirely of agents. &
Agents autonomously perform on-chain governance actions based on real-time monitoring and on-chain data analysis \\
\hline
\end{tabular}
}
\label{tab:stage-modules}
\end{table}

\subsection{Overview}

To systematically depict the evolutionary trajectory of BetaWeb, we propose a five-stage evolutionary roadmap comprising Isolated Silos, Pilot Decentralization, Assisted Execution, Hybrid Governance, and Full Autonomy, as summarized in Table \ref{tab:stage-explanation}. This categorization integrates the characteristics of blockchain and autonomous ecosystems for Agentic AI, emphasizing the progressive enhancement of system capabilities and agentic workflows. Conceptually, it parallels the five-level framework for the development of artificial general intelligence (AGI) proposed by OpenAI \citep{agiroadmap}, in which the five levels depict the evolution of AGI from passive tools to entities with capabilities that can do the work of an organization, our five-stage roadmap outlines the evolutionary path of LaMAS from passive execution to possessing autonomous collaborative and governance capabilities.

These five stages form a progressive continuum that reflects the parallel evolution of stakeholder roles, governance mechanisms, and the maturity of the business ecosystem. Figure \ref{fig:five-stage} provides a visualization of this five-stage progression, while Table \ref{tab:stage-parties} and Table \ref{tab:stage-modules} include the details of key involved parties and core system modules at each stage, respectively. Throughout this evolutionary process, the primary focus of each stage lies not only in technological improvements but also in the transforming roles of key involved parties, as well as in the shifting functional priorities and collaborative patterns of the core system modules. 

In the following five subsections, we conduct an in-depth analysis of each stage, examining their defining characteristics, major evolutionary shifts, and the strategic implications of these changes for real-world deployment and ecosystem construction. Meanwhile, we have also summarized the key technologies for each stage, which can be found in our GitHub repository\footnote{\url{https://github.com/MatZaharia/BetaWeb}}.

\subsection{Isolated Silos Stage}

The Isolated Silos stage marks the starting point of the evolutionary roadmap, corresponding to the most prevalent current state in which BetaWeb operates independently within mutually disconnected information silos. Each isolated system, like individual islands, possesses its own rules and operational pipeline, yet lacks effective bridges for data or task interaction with the external environment \citep{johnson2019no}. 

At this stage, system workflows are heavily dependent on direct human control, relying on human cognition, judgment, and actions to drive the entire process \citep{shneiderman2022human}. Agents function merely as passive tools executing fixed programs, restricted to completing predefined tasks within limited environments. Consequently, agentic workflows exhibit to be closed and linear, entirely dependent on manual triggering, intervention, and maintenance \citep{sapkota2025ai}.

In such a state, data, computational power, and intelligent services remain confined within the system’s boundaries, making cross-system resource exchange or collaboration nearly unattainable \citep{zhuang2023foundation}. While performance within a single domain may remain relatively stable, the absence of cross-domain interaction severely limits the societal value and potential for network effects. These silos not only lead to redundant construction, resource waste, and efficiency loss, but also constrain the user experience to an early-stage as ``human-driven, machine-responsive'' mode, falling short of forming an adaptive and dynamically interconnected Agentic Web \citep{sharma2025collaborative}. 

Notably, in this stage, the role of blockchain has yet to be fully realized, remaining only as a latent infrastructure layer. The key to transitioning towards the next stage lies in introducing a blockchain-based trusted identity system, enabling cross-system data verification, and establishing preliminary smart contract, thus laying the technical and governance foundations for subsequent decentralized evolution.

\subsection{Pilot Decentralization Stage}

The Pilot Decentralization stage represents the initial transition from isolated silos towards decentralized collaboration. At this stage, limited-scale interconnected channels begin to form in BetaWeb, analogous to opening certain flight routes among islands, creating small-scale trusted interconnection pilots. 

User identities undergo gradual transformation from solely managing the entire process to adopting composite roles as task requesters, execution supervisors, and agent owners \citep{borghoff2025human}. Leveraging blockchain-based identity systems, verifiable identity over agents and their actions are established. Simultaneously, agents acquire preliminary cross-domain collaboration, enabling partial intelligence sharing within a constrained decentralized network and supporting cooperative execution of cross-system tasks. Blockchain is introduced into agentic workflows as a critical trust hub for data verification and task settlement, ensuring data immutability and process traceability. However, task decomposition, allocation, and incentive mechanisms still largely depend on human negotiation and definition, with collaboration characterized by localized bilateral partnerships \citep{karim2025ai}.

This stage effectively breaks the fully isolated status by establishing trustworthy peer-to-peer cooperation, significantly reducing trust costs in cross-system interactions. Users begin to automate and indirectly manage the workflow via on-chain rules and smart contracts \citep{zou2025blocka2a}. The blockchain serve as the core technical foundation for securing cooperation and enabling initial multi-party resource integration. 

Utilizing blockchain during this stage faces challenges including cross-chain data exchange \citep{qasse2019inter}, lightweight consensus algorithms \citep{biswas2019pobt}, and scalable proofs of task execution \citep{yang2020review}. On the governance side, exploring on-chain automation for reputation evaluation, task arbitration, and incentive rules is essential \citep{he2022blockchain} but lacking. In brief, the core difficulty lies in jointly enhancing performance and security while preventing excessive ecosystem fragmentation, thus enabling steady progression towards higher levels.

\subsection{Assisted Execution Stage}

In the Assisted Execution stage, BetaWeb further evolves as inter-system connectivity resembles established regular routes and trade alliances between islands, enabling efficient circulation of information exchange and resource transactions within these networks. 

The roles of users shift towards strategic command and goal setting, with minimal intervention in execution details. Agents possess internalized domain-specific expertise and autonomously lead the operation of sub-tasks, dynamically optimizing workflows and task allocation in real-time based on task context and smart contracts \citep{zheng2025gasagent}. Agentic workflows here increasingly rely on blockchain-driven management modules. The multidimensional lifecycles of agents and tasks are fully coordinated by on-chain rules, supporting automatic invocation of off-chain computational and storage resources, thus achieving efficient on-chain and off-chain collaboration \citep{yang2024swarm}. Although automation is greatly enhanced, humans remain responsible for manual monitoring and auditing, and is required to adjust rules governed by smart contracts based on real-time system feedback \citep{10915552}.

This stage significantly improves task execution efficiency and system throughput, transforming users from executors to supervisors and effectively freeing human labor. Blockchain functions not only as a trusted foundation but also as the core platform for task scheduling and incentive distribution, promoting economic cooperation among multiple agents and forming a highly autonomous collaborative economic community. 

Multi-agent coordinated planning \citep{fang2024coordinated}, dynamic incentive model \citep{xing2024incentive}, and secure proofs for hybrid on-chain and off-chain computation \citep{chiedu2025survey} are important here. Additionally, designing on-chain economic rules with adaptive and evolutionary capabilities is essential to address the increasing task complexity and agent capabilities in a dynamic environment \citep{nguyen2024blockchain}. However, currently, agents mainly autonomously handle task execution and upstream processes, while human input for downstream monitoring, verification, and auditing remains significant. Further reducing human intervention in these areas is a focal point for future optimization.

\subsection{Hybrid Governance Stage}

In the Hybrid Governance stage, BetaWeb will establish large-scale decentralized systems spanning multiple industries and domains, with a balanced power structure among users, agents, and agentic workflows. The system possesses on-chain autonomous governance capabilities while retaining human intervention rights for critical strategic decisions, forming a governance pattern driven by both technology and institutional frameworks \citep{chaffer2024decentralized}.

Agents not only autonomously execute tasks but also actively participate in governance processes, including rule proposals and parameter adjustments through on-chain governance activities \citep{valiente2024web3}. When execution results significantly deviate from expectations, humans can intervene immediately to ensure effective control. The workflows achieve full-chain autonomy coupled with real-time supervision \citep{edwards2025human}. For example, the rules management module, driven jointly by on-chain voting and smart contracts, dynamically optimizes economic models and incentive mechanisms based on actual operation, ensuring the fairness and efficiency in governance.

The core evolution of this stage lies in a qualitative leap in governance capability, achieving a dynamic balance between autonomy and oversight. Humans primarily handle strategic formulation and critical supervision, while agents continuously improve execution and develop preliminary autonomous governance abilities, collaboratively addressing highly uncertain and complex operational environments \citep{zhuge2024agent,zhang2025webpilot}. This governance mode provides stable institutional guarantees and controllability for large-scale distributed collaboration \citep{karim2025ai}, significantly enhancing system resilience and social acceptance \citep{guerra2025iot}, promoting ecosystem expansion across industries \citep{jovanovic2022managing}, and laying a solid institutional foundation for the eventual transition to full autonomy.

This stage requires breakthroughs in constructing frameworks for continuous learning and self-evolution of agents \citep{wang2025ragen}, implementing low-latency cross-domain consensus mechanisms \citep{liu2023secure}, and designing scalable and efficient on-chain governance frameworks \citep{reijers2021now}. Additionally, exploring the legal and ethical boundaries of human-machine co-governance, building fault-tolerance mechanisms, and rapid recovery solutions are essential to ensure the sustained system operation in the face of global emergencies \citep{hu2025decentralized,de2025open}.

\subsection{Full Autonomy Stage}

Entering the ultimate stage of Full Autonomy, BetaWeb evolves into a self-regulating global intelligent society with fully autonomous capabilities. Routine task execution no longer depends on direct human intervention. Users only need to express high-level intents or set value goals, enabling the system to automatically allocate and execute tasks within a comprehensive framework of adaptive rules and incentive mechanisms \citep{chen2024s}.

Here, agents possess demand-aware capabilities for self-upgrading and self-exploring \citep{zheng2025skillweaver}, independently completing complex tasks and related social behaviors \citep{luo2025large}. They can even proactively propose new rules and adjust existing parameters on-chain. Agentic workflows achieve full automation, with task management, execution, incentive distribution, and governance all driven by blockchain, thereby constructing a closed-loop, dynamic, and boundaryless agentic collaborative ecosystem \citep{mukherjee2025agentic}.

This stage marks the transition of BetaWeb into a genuine autonomy era, where humans are fully liberated from operational or managerial roles and instead act as guides for overarching values and objectives. Through the collaboration among agents, the system attains self-governance across industries and regions. Blockchain serves as the sole foundation for global trust and resource coordination, and even driving fundamental transformations in economic and social patterns \citep{tallam2025autonomous,jaggavarapu2025evolution,wang2025internet}.

However, full autonomy also brings numerous ethical, legal, and security issues, such as value alignment, unification of cross-cultural norms, and the profound impact of self-governance systems on human societal structures \citep{hu2025trustless,chaffer2025can}. On the technical front, it is essential to guarantee blockchain’s high performance and adaptability on a global scale, while establishing verifiable self-evolution mechanisms and robust security protections \citep{fang2025comprehensive}. Addressing these challenges will determine the feasibility, reliability, and sustainable development of fully autonomous BetaWeb.

\section{Existing Products}

Currently, products or applications combining blockchain with LaMAS exhibit a diverse development landscape, spanning broad explorations from DAOs to industry-specific LaMAS. These existing solutions continuously explore new modes of collaboration among agents, aiming to leverage blockchain to enhance system transparency, trustworthiness, and security. However, the overall market remains in an early yet rapidly evolving stage, with technologies and applications not yet fully matured, and widespread adoption facing numerous challenges.

Specifically, within the Web3-oriented innovative attempts, many projects focus on building ecosystems for collaboration and interaction among users. For example, platforms including HajimeAI \citep{hajimeai}, ChainOpera AI \citep{chainoperaai}, and AGNTCY \citep{agntcy} commonly integrate smart contracts to achieve agent identity management and task scheduling, creating interoperable LaMAS that foster effective task execution environments. These projects predominantly rely on public blockchains and open-source communities, emphasizing values of openness, transparency, and decentralization. Overall, such innovations highlight open ecosystems and value co-creation, but still require breakthroughs in commercialization and large-scale impact.

In applications guided by government regulation and industry compliance, more products focus on meeting regulatory requirements and privacy protection, especially within critical sectors such as finance, healthcare, and public services. These products, such as Legion \citep{tfdlegion} and BOP \citep{Xinghuo}, typically leverage blockchain to ensure traceability and immutability of agent interactions, enabling regulators to dynamically monitor agent behavior and manage risks. Meanwhile, agent capability evaluation and incentive mechanisms are strictly incorporated within compliance frameworks to achieve secure and trustworthy applications. These efforts are mostly localized pilots tightly aligned with specific industry demands, prioritizing security and compliance. Although their application scope remains limited, such products play a vital role in safeguarding system security and enhancing industry trust, showing potential for scalable deployment under controlled environments.

In summary, based on the evolutionary roadmap presented in this paper, existing products largely remain at Stage 2. Current technological and product explorations still face the complex challenge of balancing high concurrency processing, strong trust guarantees, and data privacy protection. The triple demands have yet to be fully reconciled. This fundamental tension constitutes both a developmental bottleneck and a critical focus for future breakthroughs, underscoring the broad application prospects and profound strategic value of BetaWeb via the blockchain paradigm.

\section{Challenges and Open Problems}


In the systematic analysis of phased evolution and existing products, BetaWeb demonstrates substantial potential. Within this ambitious vision, it is crucial to proactively identify the multiple challenges and open questions that may arise during its development, which is an interdisciplinary puzzle. To provide a comprehensive assessment, we examine these issues across technological, ecosystem, and societal dimensions, aiming to elucidate current bottlenecks and outline directions for future research.

\subsection{Technological Dimension}

The construction of highly autonomous systems in BetaWeb faces multiple technical difficulties. First, the efficiency and security of cross-chain and multi-domain data exchange require significant breakthroughs. Achieving high-performance real-time collaboration while ensuring verifiability of task execution and computational trust remains a core challenge \citep{xie2025enhanced,yang2025asyncsc}. Ideally, to minimize trust costs, nearly all processes should operate on-chain, yet achieving large-scale on-chain deployment without compromising system performance remains an open problem. Second, multi-agent dynamic planning and task allocation algorithms must balance scalability and adaptability, ensuring stability and efficiency in complex and rapidly changing environments \citep{jiang2024large} while resisting potential attacks such as malicious agents, task tampering, or information asymmetry \citep{deng2025ai}. Third, blockchain plays a central role in the agentic workflows, but its performance, scalability, and low-latency communication still require optimization, particularly in global-scale applications \citep{rao2024scalability,gracy2021systematic}.

\subsection{Ecosystem Dimension}

BetaWeb involves multiple stakeholders and cross-industry cooperation, making institutional arrangements, collaborative mechanisms, and incentive design highly complex. Conflicts of interest and cooperative game problems require resolution, particularly in fully transparent and on-chain environments \citep{mohammed2025toward,popoola2024cross}. While full-chain governance can enhance transparency, their fairness may still be questioned. Issues such as last-touch attribution \citep{sriram2022return} or forward traceability \citep{mezzi2025ownsoutputbridginglaw} are required to be considered when redesigning the market mechanisms within a blockchain context to ensure equitable value distribution. Especially in agent marketplaces, questions concerning fair transaction of tasks, resources, and reputation, as well as incentives that prevent local interests from undermining overall ecosystem integrity, remain open \citep{yang2025agentexchangeshapingfuture}. Additionally, cross-domain identity verification, trust establishment, smart contract upgrades, and cross-system coordination must be systematically addressed to ensure long-term ecosystem stability and evolution \citep{ivaninskiy_are_2022}.

\subsection{Societal Dimension}

The high autonomy of BetaWeb raises a series of ethical and legal issues. Alignment of cross-cultural values and rules, responsibility attribution for autonomous agent behaviors \citep{AYAD2025100107} and potential changes in productivity and production relations \citep{Qian2024TakeIt} require careful regulation. Autonomous decisions may generate unforeseen social impacts, particularly in high-risk tasks or disputes over resource allocation, making clear responsibility assignment essential for societal trust \citep{chaffer2025decentralizedgovernanceautonomousai}. Fully transparent decision-making and transaction processes, while enhancing accountability, may also provoke new conflicts, exacerbated by the lag in legal frameworks, thereby potentially limiting widespread adoption \citep{LescrauwaetAdap}. Consequently, institutional innovation and policy guidance within a human-machine governance framework are necessary \citep{kraprayoon2025aiagentgovernancefield,VIGODAGADOT2024102530}. Furthermore, the security and resilience of highly autonomous systems pose new requirements for public safety and social stability, necessitating effective monitoring, emergency response, and recovery mechanisms \citep{9979717,Shandilya2024,jmse9060645}.

\section{Conclusion}

This paper systematically reviews the overall framework and evolutionary trajectory of BetaWeb, covering five stages from Isolated Silos to Full Autonomy. It provides a detailed exposition of the key involved parties, core system modules, and technical characteristics at each stage, alongside a survey and analysis of existing blockchain-enabled LaMAS products. The study highlights the significant role and immense potential of blockchain in realizing decentralized collaboration, trusted governance, and enhanced autonomy within LaMAS. Overall, blockchain not only offers fundamental trust mechanisms and a trustless collaborative environment for LaMAS but also facilitates the realization of autonomous governance, thereby contributing to the construction of a more open, secure, and efficient agent ecosystem.

It is noteworthy that, although this paper focuses on the driving role of blockchain technology in BetaWeb, the agents themselves also reciprocally promote the development of blockchain technology. As agent autonomy continuously advances, especially in the stages of Hybrid Governance and Full Autonomy, agents increasingly engage in on-chain governance activities such as smart contract upgrades and adaptive parameter adjustments \citep{ye2025bridging,shimony2025ai}. This high degree of autonomous agent behavior drives innovation and optimization of blockchain governance mechanisms, fostering the ongoing evolution of on-chain self-governance frameworks. Therefore, the future relationship between blockchain and BetaWeb will be characterized by bidirectional interaction and collaborative progress. Through continuous evolution, BetaWeb is expected to achieve broader practical deployment, thereby promoting comprehensive upgrades of the intelligent economy and intelligent society.

As BetaWeb continues to evolve, our future work will involve developing a prototype to test its practical viability. By conducting experiments and analyzing the results, we aim to further validate the effectiveness of blockchain-driven decentralized governance and autonomy in real-world applications. These findings will provide valuable insights into the broader deployment of BetaWeb and its potential impact on the agent-centric economy and society.

\bibliography{main}
\bibliographystyle{rlc}

\end{document}